\documentclass[12pt]{iopart}

\usepackage{graphicx}
\usepackage{float}
\usepackage{hyperref}
\usepackage{bm}  
\usepackage{siunitx}
\usepackage{overpic}

\eqnobysec

\newcommand{\CEPO}{C3PO}
\newcommand{\LIUQE}{\textsc{Liuqe}}
\newcommand{\LUKE}{\textsc{Luke}}
\newcommand{\SPECE}{\textsc{Spece}}
\newcommand{\YODA}{\textsc{Yoda}}

\newcommand{\dd}{\mathrm{d}}

\newcommand{\eqref}[1]{(\ref{#1})}

\begin{document}

\title[Suprathermal electrons in TCV]{Experimental and numerical investigation of 
suprathermal electron dynamics using vertical electron cyclotron emission}

\author{L Votta$^{1}$, M Hoppe$^{1}$, J Decker$^{2}$, E Devlaminck$^{2}$, A S Tema Biwolé$^{3}$, L Porte$^{2}$, J Cazabonne$^{4}$, Y Peysson$^{4}$ and the TCV team$^{\dag}$}
\address{$^{1}$Department of Electrical Engineering, KTH Royal Institute of Technology, SE-100 44 Stockholm, Sweden.}
\address{$^{2}$Ecole Polytechnique Fédérale de Lausanne (EPFL), Swiss Plasma Center (SPC), CH-1015 Lausanne, Switzerland.}
\address{$^{3}$Massachusetts Institute of Technology (MIT), Cambridge, MA 02139, USA.}
\address{$^{4}$CEA, IRFM, F-13108 Saint Paul-lez-Durance, France}
\address{$^{\dag}$ See author list of B. P. Duval {\em et al} 2024 Nucl.\ Fusion {\bf 64} 112023}

\ead{votta@kth.se}
\vspace{10pt}
\begin{indented}
\item[]June 2025
\end{indented}

\begin{abstract}
The Tokamak à Configuration Variable (TCV) is equipped with an advanced set of diagnostics for studying suprathermal electron dynamics. Among these, the vertical electron cyclotron emission (VECE) diagnostic offers valuable  insights into the electron energy distribution by measuring electron cyclotron emission (ECE) along a vertical line-of-sight. However, reconstructing the electron distribution from ECE measurements is inherently challenging due to harmonic overlap and thermal radiation noise. A more practical approach leverages forward modeling of ECE based on kinetic simulations. To this end, we introduce \YODA, a novel synthetic ECE diagnostic framework that simulates emission and (re)absorption of electron cyclotron radiation for  arbitrary electron distributions and antenna geometries. The framework is validated against the well-established synthetic ECE code \SPECE, using an ohmic TCV discharge as a reference case. In this study, the 3D bounce-averaged Fokker-Planck code \LUKE{} is used to model electron distributions in two electron cyclotron current drive (ECCD) experiments. The synthetic spectra generated using the combined \LUKE-\YODA{} framework successfully reproduce the main features of the experimental VECE measurements in both simulated discharges. The combination of kinetic and synthetic ECE simulations allow the identification of the features in the electron distribution function which give rise to certain signatures in the VECE signal.

\end{abstract}

\submitto{\PPCF}

\section{Introduction}
\label{sec:intro}
In magnetic confinement fusion devices such as tokamaks, suprathermal electrons can be  generated through electron cyclotron resonance heating (ECRH) and current drive (ECCD)  \cite{praterHeatingCurrentDrive2004} as a result of spatially localized wave-particle interactions,  which can significantly impact plasma heating and current drive performance. Diagnosing suprathermal electrons is challenging due to their low densities and broad energy distributions,  which span from the thermal to the relativistic regimes. Hard X-ray spectroscopy (HXRS) measurements have long provided tomographic information about suprathermal electron spatial profiles but are limited in energy and temporal resolution \cite{Choi2019SuprathermalED,Coda_2008_diagnostics}. In addition to HXRS, the electron cyclotron emission (ECE) arising from the electron gyromotion around magnetic field lines can provide information about non-thermal electrons that is better resolved in time and energy.
However, obtaining quantitative information about the non-thermal electron population can be challenging when using a horizontal line of sight (LOS) \cite{Blanchard2002,Klimanov2005}. In fact, the variation of the magnetic field strength along the LOS changes the ECE frequency in the same way as a change in suprathermal electron  energy would, and so one cannot distinguish between a change in energy and a change in  spatial location.

To overcome this issue, in the 1980s and in the 1990s, work on PLT \cite{Luce1987}, Alcator C \cite{kato1986diagnosis}, Tore Supra \cite{giruzzi1991observation,giruzzi1995measurement}, DIII-D \cite{janz1992analysis}, as well as more recently on TCV \cite{TemaBiwole2023,Arsene2024}, used a vertical ECE (VECE) configuration to study non-thermal electrons. A VECE diagnostic, using a vertical LOS, allows for a more direct determination of electron energies, as the ECE spectrum broadening due to the magnetic field gradient can be effectively reduced. Since the tokamak magnetic field varies as $\sim 1/R$, if one fixes the radial position $R$, the electron cyclotron frequency $f_\mathrm{ece}$ in the weakly relativistic limit is proportional to $n/\gamma$, where $n$ is the harmonic number while $\gamma$ is the relativistic factor. In fact, when the radiation can be attributed to a single harmonic, and within uncertainty limits arising from the VECE beam size and possible reflections, any broadening in frequency can be attributed to energy broadening alone. This allows a direct calculation of the electron kinetic energy $\mathcal{E}$ from the frequency of the emitted radiation as $\mathcal{E}=m_ec^2(\gamma-1)$, with $m_e$ the electron rest mass and $c$ the speed of light.

Although the VECE configuration provides several benefits, it is not without limitations. For example, the emission from non-thermal electrons can compete with thermal radiation from different spatial locations in the plasma at the same frequency, especially if the LOS is not isolated from wall-reflected radiation. To mitigate this issue, retro-reflectors or microwave beam dumps can be used at the termination of the LOS. Isolation of the LOS can however still be challenging if the refraction in the plasma shifts the LOS out of the dump or retro-reflector. Because of this, it may be necessary to limit the operating density and frequency to limit the effect of refraction \cite{TemaBiwole2023}. Due to these intrinsic limitations, reconstructing the electron energy distribution from ECE measurements is an ill-conditioned problem, making the diagnostic better suited for comparison against forward modelling of the electron distribution. Previous works on the Joint European Torus (JET) \cite{FiginiSPECE} rely on modelling the non-thermal ECE using the ECE synthetic diagnostic SPECE \cite{FiginiSPECE} in which sums of drifting maxwellian distribution functions are used to model suprathermal distributions. At ASDEX Upgrade (AUG), integrated modelling has been employed to calculate ECE spectra~\cite{DenkECRAD}, and recently a similar numerical analysis has been performed in \cite{Yu2024}.

In the present work, we introduce the synthetic ECE diagnostic framework \YODA{} and apply it to simulate two discharges with suprathermal electrons in TCV. Relying on this tool, we are able to calculate the  spectral ECE intensity by modelling the electron cyclotron emission and (re-)absorption for an arbitrary numerical electron distribution function, thereby allowing the possibility of accounting for effects that cannot be satisfactorily described by a sum of shifting Maxwellians, such as resonant wave-particle interaction. To calculate the electron distribution function, we use the 3-D bounce-averaged relativistic Fokker-Planck code \LUKE~\cite{LUKE3Dkinetic}. The LOS geometry of the VECE diagnostic is evaluated using the raytracing code \textsc{c3po} \cite{Peysson_2012}, which is also independently coupled with \LUKE{} to calculate the propagation of EC beams used for heating and current-drive and  model the non-inductive plasma heating in Fokker-Planck simulations.

The outline of this paper is as follows: in section \ref{sec:experiments} we describe the experimental setup and the TCV discharges used in this work. The theory of ECE along with the validation of the \YODA{} synthetic diagnostic framework are described in section \ref{sec:modelling}. Finally, in section  \ref{sec:results}, we compare Fokker-Planck and synthetic diagnostic simulations to experimental data, before concluding in section \ref{sec:conclusions}.

\section{Experimental setup}
\label{sec:experiments}
The experiments analyzed in the present work are conducted on TCV~\cite{Duval2024} (major radius $R=\SI{0.89}{m}$, minor radius $a  = \SI{0.25}{m}$, magnetic field  $B_0 = \SI{1.4}{T}$). In TCV, profiles of electron temperature $T_e$ and density $n_e$ are measured using Thomson scattering (TS). While the spatial resolution of TS is high, it offers a  relatively low temporal resolution, approximately $\SI{60}{Hz}$, achieved by using three pulsed lasers, each operating at a repetition rate of $\SI{20}{Hz}$ \cite{Blanchard_2019}. Additionally, magnetic measurements are employed to determine the plasma current $I_{\rm p}$ and loop voltage $V_{\rm loop}$. These, together with TS measurements and magnetic coil current prescriptions, can be fed to the magnetic equilibrium reconstruction code \LIUQE~\cite{Moret2015} to yield axisymmetric magnetic equilibria, which are utilized in our simulations.

To produce suprathermal electrons, the TCV ECRH system~\cite{Paley2009} was used. The system consists of five gyrotrons, two of which operate at a frequency of $\SI{82.7}{GHz}$ with a nominal power output of $\SI{680}{kW}$, and three of which operate at $\SI{118}{GHz}$ with a nominal power output of $\SI{480}{kW}$. Additionally, there are two dual frequency gyrotrons with a nominal power output of $\SI{1000}{kW}$, capable of heating at either $\SI{84}{GHz}$ or $\SI{126}{GHz}$. The plasma heating provided by the gyrotrons is achieved through the cyclotron resonance at either the second (X2) or the third (X3) harmonic of the cyclotron frequency. The EC waves are injected into the plasma through a launcher capable of dynamically changing the EC injection angle $\theta_l$, allowing resonances in different regions of the electron momentum space.

\subsection{Vertical electron cyclotron emission}\label{sec:experiments:vece}

\begin{figure}
    \centering
    \includegraphics[width=.49\textwidth]{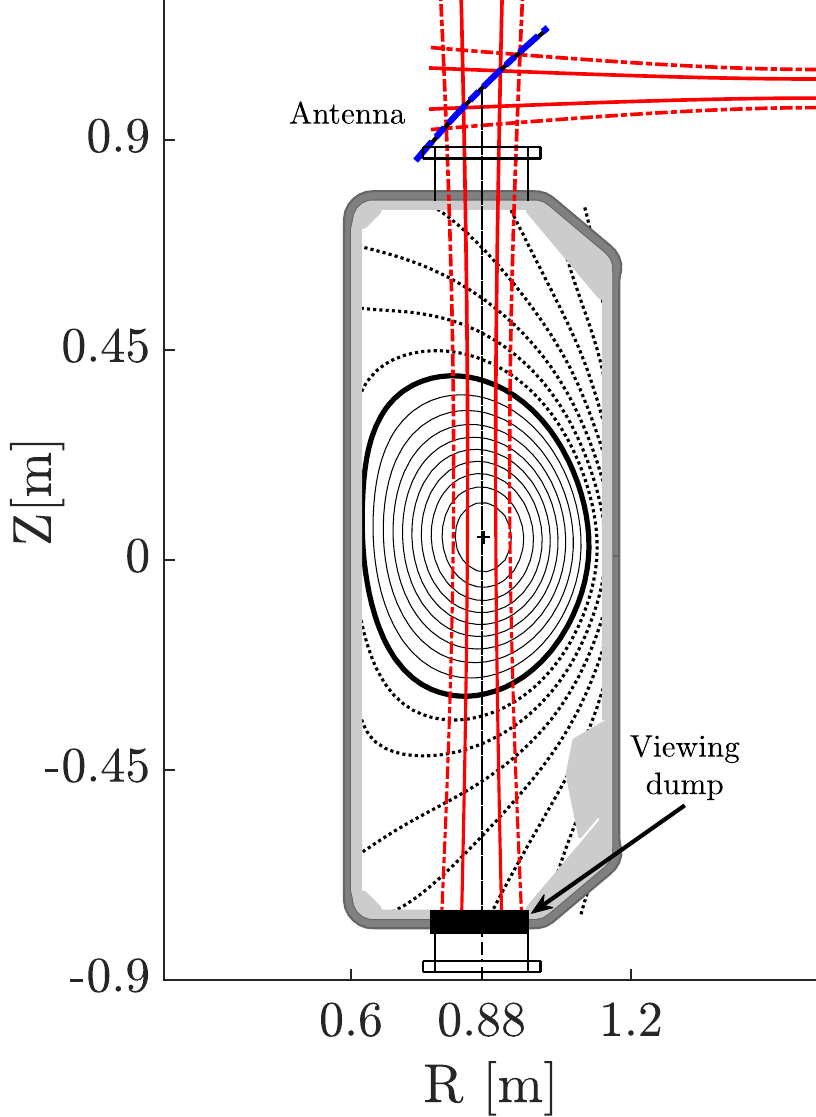}
    \caption{Poloidal view of the VECE diagnostic lines-of-sight in TCV.}
    \label{fig:VECELOS}
\end{figure}

The diagnostic of primary interest is the vertical ECE (VECE) system on TCV; its line-of-sight geometry is shown in Fig.~\ref{fig:VECELOS} \cite{TemaBiwole2023,TemaBiwolePhD}. The diagnostic measures the emission in the \qtyrange{78}{114}{GHz} band with \SI{10}{\micro\second} temporal resolution and \SI{750}{MHz} channel bandwidth. A wire-grid polarizer separates the X- and O-mode components prior to detection.

The optical design of the diagnostic ensures that the vertical antenna pattern forms a Gaussian beam with a waist radius of approximately \SI{3}{cm} at the mid-plane of the tokamak. In vacuum conditions, the beam terminates on a highly absorbing viewing dump located at the bottom of the vessel. As shown in Ref.~\cite{ArseneViewingDump}, this viewing dump effectively suppresses reflections and ensures that the detected radiation originates from within the field-of-view, greatly simplifying modelling.

The diagnostic detects fast electron emission from the downshifted third or fourth harmonics. Since the frequency of the emitted radiation is inversely proportional to the electron energy, it can be estimated that the VECE diagnostic can measure radiation from electrons with energies of up to approximately $\SI{250}{keV}$ without encountering harmonic overlap~\cite{TemaBiwole2023}. Reflected thermal emission can contribute significantly when the LOS is refracted and misses the viewing dump. In such a configuration, studied extensively in Ref~\cite{Arsene2024,TemaBiwolePhD}, only VECE frequencies for which thermal ECE emission is negligible everywhere in the plasma are exploitable, and the toroidal magnetic field may be adjusted to vary the selection of such eligible frequencies.

\subsection{Experiments}
In section~\ref{sec:results}, we will consider two different ECRH scenarios in TCV previously analyzed in Ref.~\cite{TemaBiwolePhD,Arsene2024}, which illustrate two different techniques of modelling the fast electrons. In the first case, the EC launcher was held at a fixed angle for the duration of its activation in order to generate stable plasma conditions for studying fast electrons. In the second case, the launcher angle was instead swept in discrete steps across five different angles in order to study the effect of the launcher angle on the fast electron distribution function.

\subsubsection{Constant ECCD launching angle}
\begin{figure}
    \centering
    \begin{overpic}[width=\textwidth]{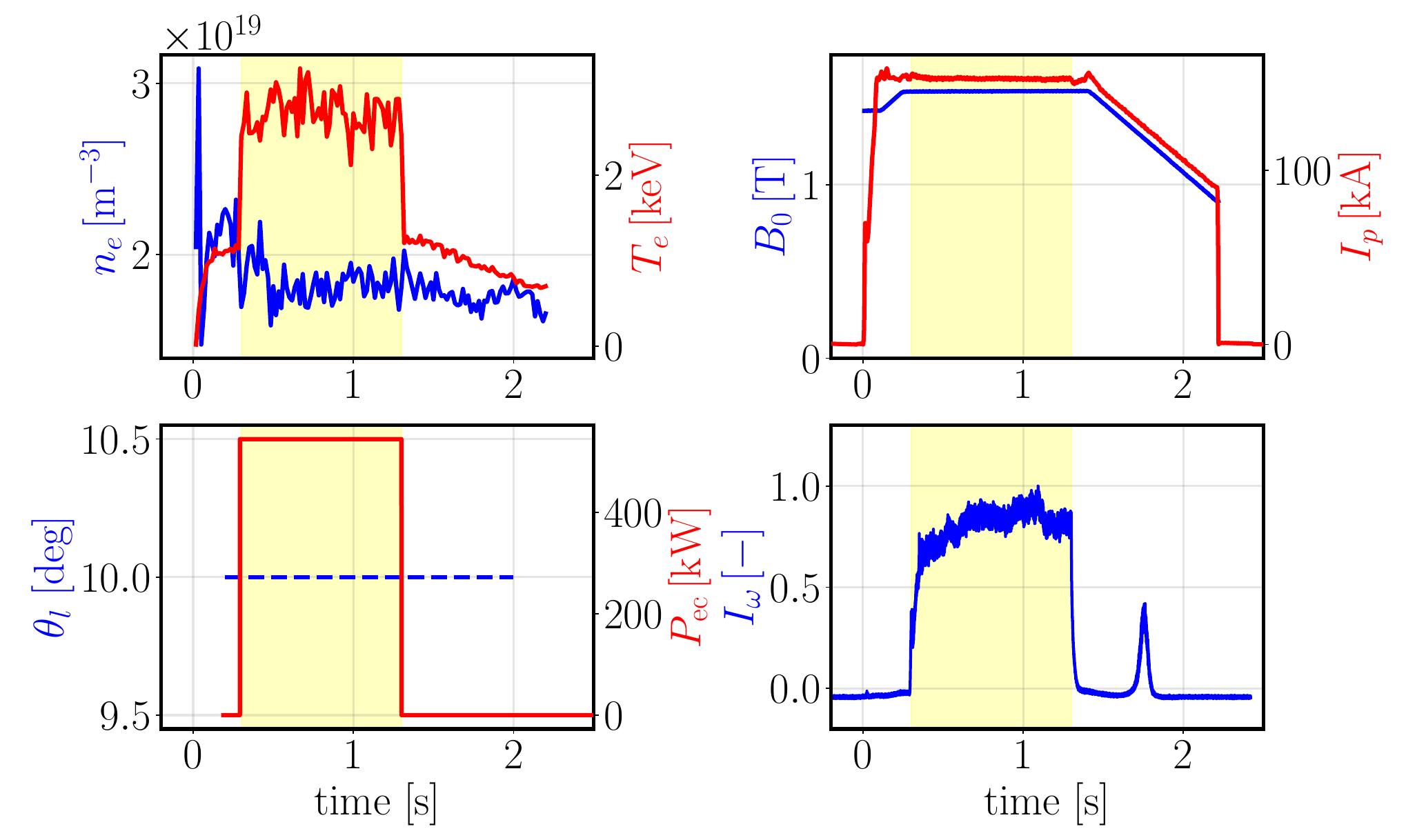}
        \put(42,58){\textbf{TCV \#73127}}
        \put(37,52){(a)}
        \put(84,52){(b)}
        \put(37,26){(c)}
        \put(59,26){\qty{108.84}{GHz}}
        \put(84,26){(d)}
    \end{overpic}
    \caption{
        Experimental plasma parameters in discharge \#73217. Electron cyclotron heating is enabled for the duration indicated by the shaded area, between $t=\qtyrange{0.3}{1.3}{s}$. (a) Electron density (blue, left axis) and temperature (red, right axis) from Thomson scattering. (b) On-axis magnetic field strength $B_0$ (blue, left axis) and plasma current $I_{\rm p}$ (red, right axis). (c) ECH launcher angle (dashed blue, left axis) and injected ECH power (red, right axis). (d) Measured VECE intensity at a frequency $\qty{108.84}{GHz}$.
    }
    \label{fig:exp73217}
\end{figure}

The TCV plasma discharge \#73217 was specifically designed to achieve VECE measurements in which both thermal and non-thermal EC emission could be observed separately during the discharge. As indicated by the shaded region in figure~\ref{fig:exp73217}, EC heating was enabled for about one second between \qty{0.3}{s} and \qty{1.3}{s} at a constant on-axis magnetic field strength of \qty{1.54}{T}. The magnetic field strength was kept constant for \qty{100}{ms} after the heating was turned off until $t=\qty{1.4}{s}$, at which point it was ramped down from \qty{1.54}{T} to \qty{0.9}{T}.

During the heating phase, the ECH power was held constant at \qty{500}{kW}, and launched horizontally in co-current configuration such that the parallel index of refraction is $N_\parallel=-0.2$ in the power deposition region. In this discharge, the plasma current was ramped down along with the magnetic field to maintain the flux surface geometry. This preserved the X-mode polarization from the current drive to the calibration phase.

An interesting feature of \#73217 is the smooth ramp-down of the magnetic field after $t = \qty{1.4}{s}$, designed to identify the thermal ECE intensity peaks at each frequency as the corresponding X3 resonance crosses the beam path~\cite{Arsene2024}. Figure~\ref{fig:exp73217}d shows how, for the measured frequency of \qty{108.84}{GHz}, the thermal peak is observed around $t=\qty{1.7}{s}$, corresponding to an on-axis magnetic field strength of $B_0\approx\qty{1.2}{T}$. As is demonstrated in section~\ref{sec:results}, the variation of the thermal peak frequency enables calibration of the diagnostic by matching the measured VECE channel intensity to the calculated synthetic ECE intensity at each resonance.

The radiation intensity decreases in all VECE channels as soon as the heating is turned off at $t=\qty{1.3}{s}$, thereby indicating that the radiation detected during the current drive phase is mainly produced by non-thermal electrons, with little or no contribution from thermal electrons at the observed frequency \qty{108.84}{GHz}. The intensity level of the polluting background radiation is higher for lower frequencies due to the X2 cold resonances of lower frequencies being located at higher major radii in the plasma. In fact, for frequencies that have their thermal X2 cold resonance position in the plasma, the measured radiation during current drive is a combination of both thermal and non-thermal radiation. In the specific case of interest, the chosen frequency of \qty{108.84}{GHz} is sufficiently high that the background radiation is not polluting the measurements.

\subsubsection{Varying ECCD launching angle}
\begin{figure}
    \centering
    \begin{overpic}[width=\textwidth]{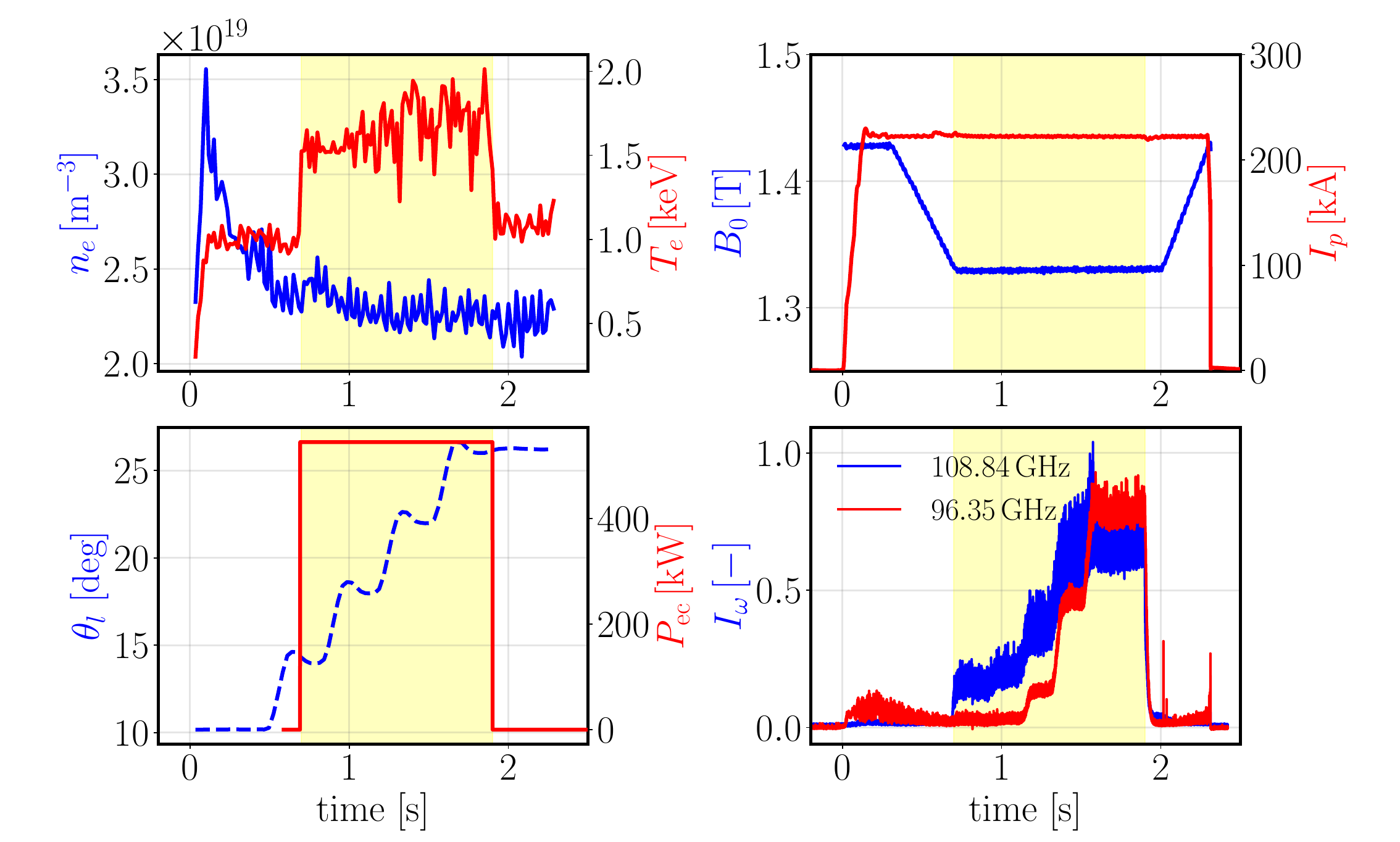}
        \put(40,59){\textbf{TCV \#72644}}
        \put(38,53){(a)}
        \put(84,53){(b)}
        \put(38,26){(c)}
        \put(84,26){(d)}
    \end{overpic}
    \caption{
        Experimental plasma parameters in discharge \#72644. Electron cyclotron heating is enabled for the duration indicated by the shaded area, between $t=\qtyrange{0.7}{1.9}{s}$.
        (a) Electron density (blue, left axis) and temperature (red, right axis) from Thomson scattering. (b) On-axis magnetic field strength $B_0$ (blue, left axis) and plasma current $I_{\rm p}$ (red, right axis). (c) ECH launcher angle (dashed blue, left axis) and injected ECH power (red, right axis). (d) Measured VECE intensity at the frequencies \qty{108.84}{GHz} (blue) and \qty{96.35}{GHz} (red).
    }
    \label{fig:exp72644}
\end{figure}

In TCV discharge \#72644, shown in figure~\ref{fig:exp72644}, a total ECH power of $P_{\rm ECH}=\qty{500}{kW}$ was injected into the plasma between \qtyrange{0.7}{1.9}{s} while the launcher toroidal angle was varied in five discrete steps. The angle was varied from \qtyrange{10}{26}{\degree}, giving rise to the stair-shaped VECE signal shown in figure~\ref{fig:exp72644} for the frequencies $\qty{108.84}{GHz}$ and $\qty{96.35}{GHz}$ for which the contribution from reflections of X2 radiation becomes negligible as the magnetic field is decreased to $B = 1.34 \, \rm T$~\cite{Arsene2024}, as observed in Fig.~\ref{fig:exp72644}(b,d).

The measured VECE intensity increased sharply from noise levels as the heating was turned on at $t=\qty{0.7}{s}$. This indicates that the measurements are not polluted by background radiation, but comes exclusively from fast electrons. At the ECH onset, only the higher frequency shows a sharp increase in measured intensity. The lower frequency signal, corresponding to a higher electron energy, increases only at the third step in toroidal launching angle. Note however that the jump corresponding to the third switch of the launching angle is higher than that of the higher frequency. 

\section{Electron cyclotron emission and absorption modelling}
\label{sec:modelling}
To calculate the electron cyclotron emission (ECE) intensity at the antenna location, $I_\omega(s_{\rm ant})$, a series of steps is needed. First, we must integrate the \emph{ray-tracing equations} that govern the propagation of electron cyclotron (EC) waves along the given line-of-sight (LOS). Once the ray path is known, the \emph{radiation transport equation} is to be solved along the path to determine the continuous emission and (re-)absorption of EC waves by the plasma. Both the emission and absorption processes depend on the electron distribution function $f_e$, which is obtained by solving the Fokker-Planck equation. These are the steps which will be carried out in section~\ref{sec:results}, and in this section we will describe the action of each step in more detail. Crucially, we will introduce the \YODA{} code which evaluates the EC emission from an arbitrary numerical electron distribution function.

\subsection{Electron cyclotron wave propagation}
In quasi-stationary plasmas, for which the equilibrium scale length is much longer than the wavelength of the propagating wave, it is feasible to compute the propagation of the wave via ray-tracing. Approximating the wave as a ray necessarily means disregarding diffraction of the wave, while refraction is still accounted for. This technique has been extensively employed in the analysis of lower-hybrid and EC wave propagation \cite{Bonoli1984,bizarro1993self,peysson1996magnetic,Imbeaux2005} and will be utilized in this work to simulate the ECE antenna pattern using the \CEPO{} ray-tracing code \cite{Peysson_2012}. The \CEPO{} code solves the ray-tracing equations
\begin{equation}
    \frac{\dd\boldsymbol{r}}{\dd t} = -\left(
        \frac{\partial\Lambda/\partial\boldsymbol{k}}
        {\partial\Lambda/\partial\omega}
    \right)_{\Lambda=0},\quad
    \frac{\dd\boldsymbol{k}}{\dd t} =
        \left(
            \frac{\partial\Lambda/\partial\boldsymbol{r}}
            {\partial\Lambda/\partial\omega}
        \right)_{\Lambda=0},
\end{equation}
where $\boldsymbol{r}=\boldsymbol{r}(t)$ is the ray path, $\boldsymbol{k}=\boldsymbol{k}(t)$ is the wave vector, $\omega$ is the wave angular frequency, and $\Lambda$ is the determinant of the dispersion tensor.

In the ray-tracing approach, the antenna pattern is not considered as a whole beam, but is rather divided into a finite number of rays, each evaluated individually to produce a gaussian pattern at a given EC frequency. Although the rays propagate from the electrons to the antenna in reality, evaluating the rays in this manner is computationally inefficient, since most rays will never reach the antenna. A better approach is to utilize the reciprocity theorem of electromagnetic fields and solve for the reciprocal ray paths from the antenna into the plasma and onto the viewing dump. Thus, for the purposes of this article, a vertical synthetic EC wave launcher is defined in \CEPO{}, the rays are launched from the antenna position, and are propagated through the plasma until the last closed flux surface (LCFS) is reached. The ray trajectories in the vacuum before and after the LCFS are modeled as straight lines and the wave polarization at the plasma boundary is assumed to be conserved. Only a single passage of the rays through the plasma is considered, since \YODA{} does not model the reflection from the tokamak inner walls. This assumption of no reflection is motivated by the presence of a viewing dump in LOS of the TCV VECE system, as detailed in section~\ref{sec:experiments:vece}.

Since the magnetic field strength varies with the major radius coordinate, the EC resonance condition for a given electron and emission frequency is met at a distinct major radius. This results in a stratified emission profile, where different spatially finite layers correspond to different frequencies of emitted radiation which build the EC emission layer. The presence of this region is crucial when launching rays from the top of the plasma, as some of them may miss the emission layer, resulting in an inaccurate representation of the real intensity distribution. To address this problem, the \CEPO{} code is capable of using an arbitrary number of rays $n_{\rm rays}$ inside the antenna beam using a parametrization of the rays such that we place them on circles with $n_{\rm radial}$ circles and $n_{\rm angular}$ rays on each circle to form a Gaussian beam.

\subsection{Electron cyclotron radiation transport}
The transport of EC radiation through the plasma is described by the radiation transport equation~\cite{bekefi}
\begin{equation}
    \label{eq:transport}
    \frac{\dd}{\dd s} \left[ \frac{I_\omega(s)}{N_{\omega,r}^2(s)}\right]=\frac{1}{N_{\omega,r}^2(s)} \biggl[ j_\omega(s) -\alpha_\omega(s) I_\omega(s) \biggl].
\end{equation}
All quantities in this equation are local functions of the arc-length $s$, the coordinate of the ray trajectory through the propagation medium, while $N_{\omega,r}$ denotes the ray refractive index which can be set to one if only the intensity at the ECE antenna position is of interest \cite{Bornatici}. This assumption has also been demonstrated in Ref.~\cite{Bornatici} to be valid for non-thermal electron distributions. Equation~\eqref{eq:transport}, having the form of a power balance equation, accounts for the continuous emission and (re-)absorption of EC waves respectively through the emissivity coefficient $j_\omega$, which is the source term, and the absorption coefficient $\alpha_\omega$, representing the losses. To calculate the spectral intensity $I_\omega(s_\mathrm{ant})$ at the antenna, the radiation transport equation must be integrated along the trajectory of each individual ray in the antenna pattern, starting from the position of the viewing dump (at $s=0$), with the zero reflection boundary condition $I_\omega(0)=0$.

A suitable expression for the absorption coefficient, formulated as an integral over momentum space $(p_\parallel, p_\perp)$ with both coordinates normalized to $m_ec$, is found in the literature to be~\cite{Albajar}
\begin{equation}
    \label{eq:alphacoeff}
    \alpha_{\omega}^{(n)}(s) = -2\pi^2 \frac{\omega_{\rm p,0}^2}{c \omega} \int \int \left|\boldsymbol{\Theta}\right|^2 \hat{R}_n f_e \, \delta \left( \gamma - p_\parallel N_\parallel - \frac{n}{\bar{\omega}} \right) \frac{p_\bot}{\gamma} \, \dd p_\bot \dd p_\parallel,
\end{equation}
where the operator $\hat{R}_n=n \Omega_c /(\omega p_\perp)\partial/\partial p_\perp+N_\parallel \partial/\partial p_\parallel$. Similarly, the emissivity can be derived from Ref.~\cite{Belotti1997} and is found to be
\begin{equation}
    \label{eq:emisscoeff}
    j_{\omega}^{(n)}(s) = m_e c^2 \frac{N_{\omega,r}^2 \omega_{\rm p,0}^2 \omega}{4 \pi c^3} \int \int \left|\boldsymbol{\Theta}\right|^2 f_e \, \delta \left(\gamma - p_\parallel N_\parallel - \frac{n}{\bar{\omega}}\right) \frac{p_\bot}{\gamma} \, \mathrm{d} p_\bot \, \mathrm{d} p_\parallel,
\end{equation}
with $n$ denoting the harmonic number and $f_e(p_\parallel,p_\perp)$ the electron distribution function. In both of equations \eqref{eq:alphacoeff} and \eqref{eq:emisscoeff}, the relativistic factor is $\gamma=\sqrt{1+p^2}$, while $\bar{\omega}=\omega/\Omega_c$ and $\boldsymbol{\Theta}$ is the polarization factor which takes into account the wave polarization and finite Larmor radius effects.  The equations also contain the plasma frequency $\omega_{\rm p}$, the speed of light in vacuum $c$, and the parallel and perpendicular components of the plasma refractive index, $N_\parallel$ and $N_\perp$ respectively. Lastly, the quantity in the Dirac delta function is the relativistic EC resonance condition, which also accounts for the Doppler shift.
    
This model for the EC absorption and emission, along with a solver for the radiation transport equation, is implemented in the ECE synthetic diagnostic framework \YODA, developed for the present study. This tool distinguishes itself from other ECE synthetic diagnostic codes by its ability to calculate ECE spectra from arbitrary numerical electron distribution functions which, in non-thermal scenarios, can strongly affect the emission and (re-)absorption of EC waves by the plasma.

\subsection{Non-thermal electron distribution function}
To model the electron distribution function, we use the \LUKE{} code~\cite{LUKE3Dkinetic}. It is a fully relativistic 3D bounce-averaged Fokker-Planck solver which takes as input numerical axisymmetric equilibrium data, and is integrated with the databases of various tokamak experiments, including TCV which is studied here. In addition to solving the Fokker-Planck equation, \LUKE{} can also non-linearly solve for the electric field profile in the plasma via the induction equation. The code is formulated in a numerically conservative formalism, ensuring the numerical conservation of electron density. 

The \LUKE{} code solves the linearized bounce-averaged drift kinetic equation for electrons
\begin{equation}
    \frac{\partial f_{e}}{\partial t} =
    \sum_s \mathcal{C}\left(f_{e},f_{s0}\right) +
    \mathcal{E}\left(f_{e}\right) +
    \mathcal{Q}\left(f_{e}\right) +
    \mathcal{S}\left(f_{e}\right),
    \label{eq:DKE}
\end{equation}
where $\mathcal{C}$ denotes the collision operator, $\mathcal{E}$ denotes the electric field acceleration operator, $\mathcal{Q}$ is the quasi-linear diffusion operator which is used for modelling the interaction of electrons with EC waves and is evaluated with the ray-tracing code \CEPO, and $\mathcal{S}$ describes the radial transport of electrons. The radial transport of fast electrons is primarily believed to arise from turbulent fluctuations, which cannot be described from first principles in \LUKE, and so a general advection-diffusion operator with ad-hoc transport coefficients is used to describe the radial transport, namely
\begin{equation}
    \mathcal{S}(f_{e}) = \frac{1}{V'}\frac{\partial}{\partial r} \left[ V'\left(\mathcal{D}_r \frac{\partial f_e}{\partial r} - \mathcal{F}_r f_{e}\right)\right].
\end{equation}
Here, $\mathcal{D}_r$ and $\mathcal{F}_r$ are the diffusion and advection coefficients, respectively, and $V'$ is the configuration space jacobian. In the present work, we take $\mathcal{F}_r=0$ and ignore any spatial variation of the radial diffusion coefficient.

\subsubsection{Induction equation}
To account for the electric field response to a time-varying plasma current density $\boldsymbol{J}$, \LUKE{} can also solve the induction equation
\begin{equation}
    \label{eq:luke_induction_toroidal}
        \nabla^2 \boldsymbol{E}=\mu_0 \frac{\partial\boldsymbol{J}}{\partial t},
\end{equation}
where $\mu_0$ is the permeability of free space and $\boldsymbol{E}$ and $\boldsymbol{J}$ are the electric field and current density respectively. The electric field $\boldsymbol{E}=\nabla \Phi + \partial \boldsymbol{A}/\partial t$ results from either a non-zero electrostatic potential $\Phi$, from induction by means of a transformer, or from a time varying plasma current, described by a magnetic vector potential $\boldsymbol{A}$. The electrostatic potential arises due to transport, vanishing in the toroidal direction. The potential is typically of the order $\sim T_e/e$~\cite{helandersigmar2002}, which allows one to assume that the induced component of the electric field is purely toroidal within the assumptions of \LUKE. As shown in~\ref{AppendixA}, the induction equation then reduces to the single partial differential equation
\begin{equation}
    \label{eq:LUKEinduction}
        \frac{B_0}{\bar{q}}\frac{\partial}{\partial \psi}\left[B_0 l(\psi)\frac{\partial V_{\rm l}}{\partial \psi} \right]= 2\pi\mu_0 \frac{\partial \langle J_\phi \rangle}{\partial t}, 
\end{equation}
where $V_{\rm l}(\psi)\equiv 2\pi RE_\phi$ denotes the toroidal loop voltage which is constant on the flux surface $\psi$, and
\begin{eqnarray}
    l(\psi)&=\frac{1}{B_0}\int_0^{2\pi} \frac{\dd \theta}{2 \pi}\frac{r|\nabla\psi|}{\cos \alpha},\\
    \bar{q}(\psi)&=\int_0^{2\pi}\frac{\dd \theta}{2\pi}\frac{rB_0}{\cos\alpha |\nabla\psi|},
\end{eqnarray}
where the angle brackets in equation \eqref{eq:LUKEinduction} denote a flux surface average, as defined in \ref{AppendixA}.

\subsection{Validation of YODA}
\begin{figure}
    \centering
        \begin{overpic}[width=.5\textwidth]{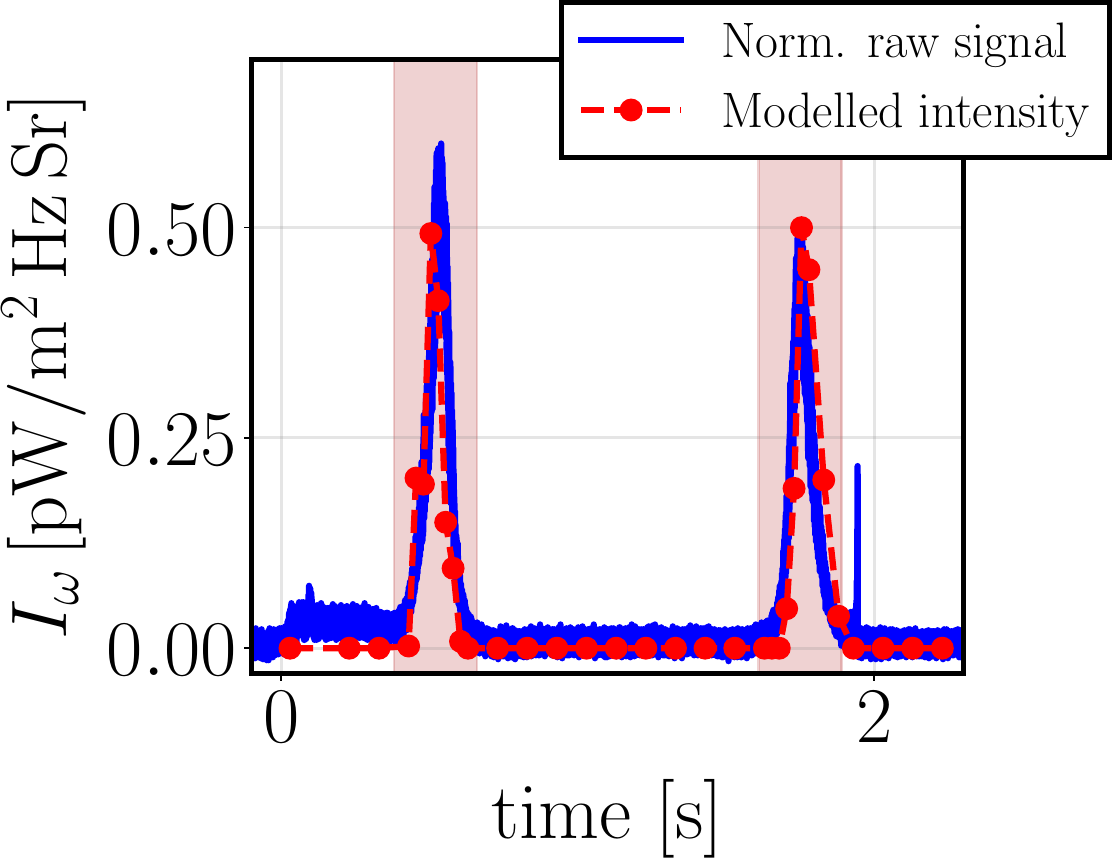}
        \put(25,66){(a)}
    \end{overpic}
            \begin{overpic}[width=.44\textwidth]{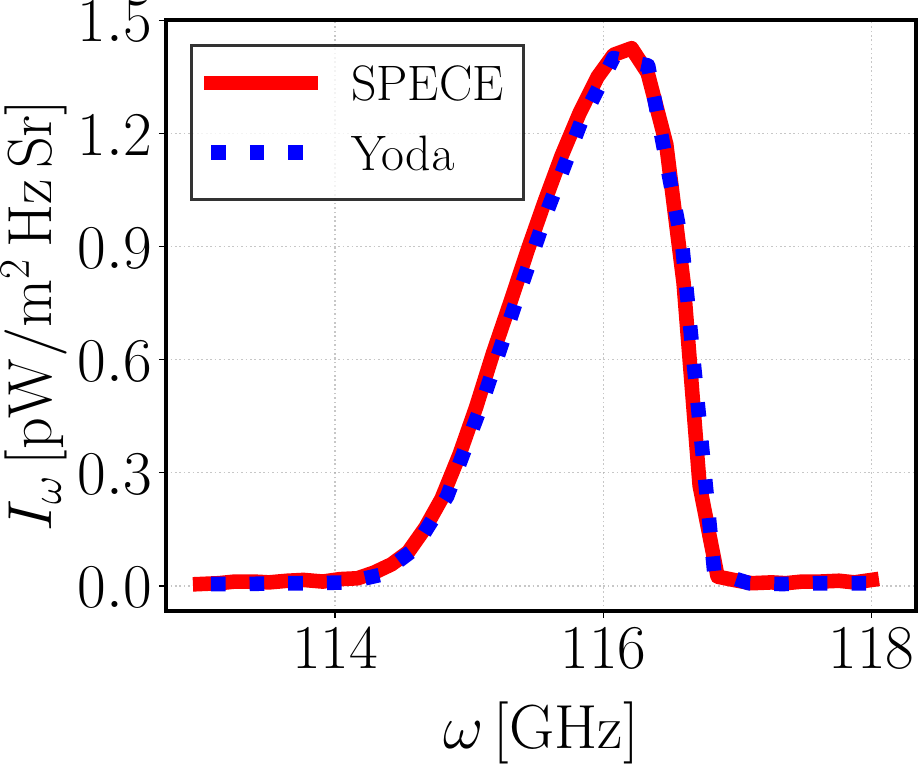}
        \put(90,74){(b)}
    \end{overpic}
            \caption{
        The synthetic ECE diagnostic \YODA{} is (a) validated by comparing simulations against experimental measurements in the thermal discharge \#73003, and (b) verified by comparing to the synthetic ECE diagnostic \SPECE.
    }
    \label{fig:yoda:vv}
\end{figure}

To validate \YODA, we simulate TCV discharge \#73003 which is free from suprathermal electrons. In this discharge, the magnetic field strength $B_0$ is varied between $\qtyrange{1.4}{0.9}{T}$. Since the frequency of the emitted cyclotron radiation depends on $B_0$, and since the VECE diagnostic observes at one major radius, the signal at a given frequency will only be non-zero at one specific value of $B_0$ and we thus expect to see a peak in the radiation within a finite time interval during the magnetic field ramp. Figure  \ref{fig:yoda:vv}a shows that the intensity predicted by \YODA{} captures the trend of the ECE thermal emission as seen along a vertical LOS at a frequency of \qty{108.84}{GHz}, which corresponds to the third harmonic extraordinary mode (X3). The slight misalignment between the modelled intensity and the two measured thermal peaks is due to the angular deflection of the
VECE antenna pattern due to mechanical vibrations.

To assess quantitatively the validity of the \YODA{} calculations, they are benchmarked against the ECE synthetic diagnostic \SPECE,  which was previously validated against TCV thermal plasmas \cite{TemaBiwole2023}. Here, we compare the calculated intensity at $t=\qty{1.9}{s}$, corresponding to a magnetic field strength for which the X3 thermal emission is in  the frequency range of \qtyrange{113}{118}{GHz}. We consider only the central ray in the antenna pattern to validate the \YODA{} calculations against \SPECE. As shown in figure \ref{fig:yoda:vv}b, the \YODA{} calculations are in good agreement with the \SPECE{} results for this test discharge.

\section{Modelling suprathermal electrons}
\label{sec:results}
This chapter discusses the numerical modelling and analysis of suprathermal electron dynamics in TCV, using the  Fokker-Planck solver \LUKE{} coupled with the synthetic ECE diagnostic \YODA{}. We consider two different scenarios which exhibit different features which are of interest in suprathermal electron studies, which were previously analyzed in Ref.~\cite{Arsene2024,TemaBiwolePhD}. In the first case, \LUKE{} is used to solve the steady-state Fokker-Planck equation in order to model a stair-shaped VECE intensity pattern, while in the second case the time-dependent Fokker-Planck equation is solved to study the collisional relaxation of the distribution function when EC heating is disabled.

\subsection{Time-asymptotic Fokker-Planck simulations}
The Fokker-Planck solver \LUKE{} can be used to solve the time-asymptotic Fokker-Planck equation. In doing so, it is assumed that the loop voltage profile remains radially uniform over a time interval which is much longer than the collision time. For TCV discharge \#72644, in which the EC toroidal angle changes six times, our goal is to describe the fast electron dynamics during steady-state ECH phases, i.e.\ when the ECH toroidal injection angle remains fixed and the injected power is constant. Under these conditions, the assumption of a uniform loop voltage holds, making the time-asymptotic approach appropriate for this analysis.

To compare the distribution functions with TCV experiments, \LUKE{} is coupled to the ECE synthetic diagnostic \YODA{} presented in section~\ref{sec:modelling}.
For \#72644, we run six time-asymptotic \LUKE{} simulations, one for each ECH toroidal injection angle. Each simulation starts from a Maxwellian distribution function, and the electric field is estimated from the loop voltage $V_\mathrm{loop}$ on the machine inner wall. Radial electron transport is included in the Fokker-Planck equation via a radial diffusion operator, for which we choose the diffusion coefficient
\begin{eqnarray}
    \mathcal{D}_r &= D_{r,0} \, \mathcal{H}\left( p_\parallel - p_{\parallel,\mathrm{min}}\right),
    \label{eq:ESmodel}
\end{eqnarray}
where $\mathcal{H}(p)$ is the Heaviside step function, $p_{\parallel,\mathrm{min}} = 3 p_\mathrm{th}$ is the lower bound for the transport prescribed by the code to allow the conservation 
of the number of electrons, and $D_{r,0}$ is a free parameter tuned so that the simulated plasma current matches experimental measurements. A value of $D_{r,0}=\qty{3}{m^2/s}$ yields good agreement with the experimentally measured plasma current.

\begin{figure}
    \centering
        \begin{overpic}[width=.49\textwidth]{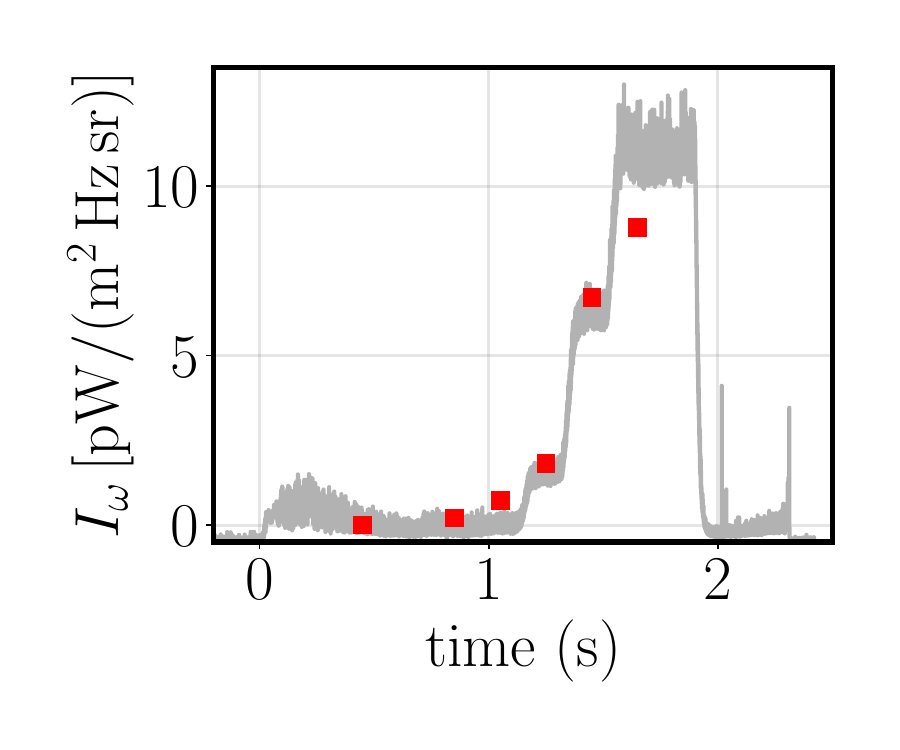}
        \put(22,78){(a) \SI{90}{keV}}
    \end{overpic}
        \begin{overpic}[width=.49\textwidth]{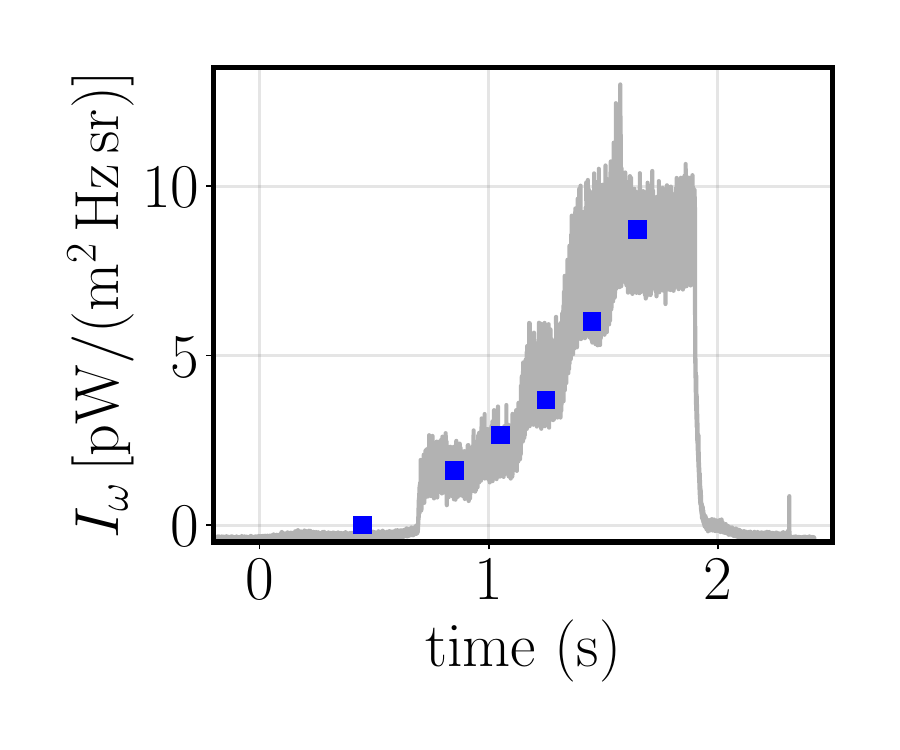}
        \put(22,78){(b) \SI{20}{keV}}
    \end{overpic}
    \caption{
        Comparison between VECE normalized raw measurements and the modelled intensity (3rd harmonic X-mode) using an ad-hoc radial transport operator at (a) $f_\mathrm{ece}\sim\SI{96}{GHz}$ ($\mathcal{E} \sim\SI{90}{keV}$) and (b) $f_\mathrm{ece}\sim\SI{109}{GHz}$ ($\mathcal{E} \sim\SI{20}{keV}$), respectively.
    }
    \label{fig:VECE_synthetic_72644}
\end{figure}

Figure~\ref{fig:VECE_synthetic_72644} shows the experimental and simulated VECE signals as functions of time. As the toroidal angle of the ECH launcher is swept in discrete steps, the VECE signal increases in corresponding steps.
While the simulations capture the overall trend that an increasing toroidal angle causes a stronger VECE signal, the exact relative difference between the different VECE levels is not fully reproduced.

\begin{figure}
    \centering
    \includegraphics[width=0.8\linewidth]{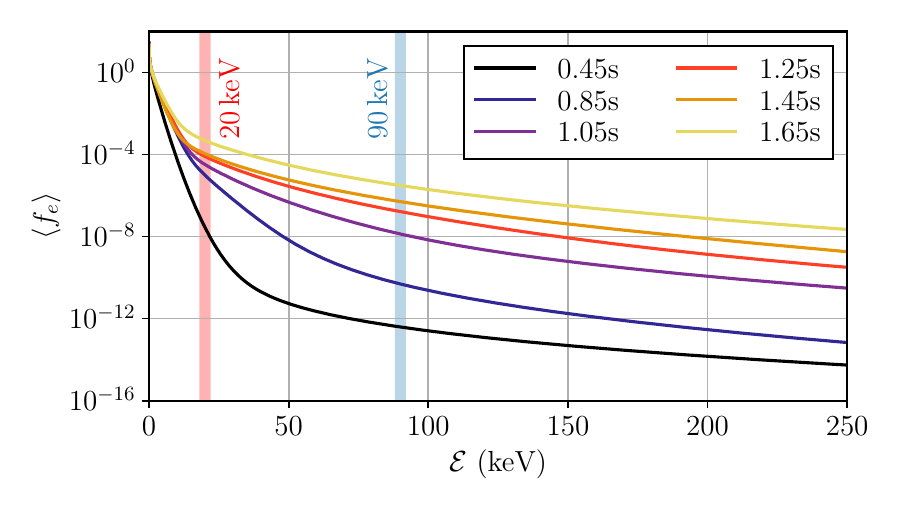}
    \caption{Evolution of the electron energy spectrum near the magnetic axis in discharge \#72644. Each time point corresponds to a step in the staircase pattern shown in figure~\ref{fig:VECE_synthetic_72644}. The evolution of the distribution function at other radii is nearly identical to this.}
    \label{fig:72644:f}
\end{figure}

The time evolution of the distribution function in \#72644 is displayed in figure~\ref{fig:72644:f}, with shaded regions corresponding to the energies which the \qty{96}{GHz} and \qty{109}{GHz} channels of the VECE should be sensitive to (corresponding to \qty{90}{keV} and \qty{20}{keV} electron kinetic energy respectively). The figure illustrates how the higher frequency channel should observe close to the thermal bulk, and thus sense suprathermal electrons as soon as they escape the bulk, while the lower frequency channel observes electrons further out in the tail.

The most significant differences in the two VECE signals is that the \qty{96}{GHz} (\qty{90}{keV}) channel primarily rises in the last three time points, while the \qty{109}{GHz} (\qty{20}{keV}) channel rises in relatively even steps throughout the discharge. The reason behind this can be understood from the distribution functions in figure~\ref{fig:72644:f}: at \qty{20}{keV}, the distribution remains close to Maxwellian before ECH is enabled. As soon as the ECH is enabled, the temperature increases and a tail of electrons is pulled out, leading to a drastic increase in the electron density at \qty{20}{keV}. Subsequent changes of the EC toroidal angle only leads to slight changes in the magnitude of the electron tail, due to the proximity of this energy to the region where collisions dominate and the distribution is therefore Maxwellian. While we only show the distribution function close to the magnetic axis in figure~\ref{fig:72644:f}, the evolution is nearly identical at other radii.

At \qty{90}{keV}, the electrons are significantly less collisional and well separated from the thermal bulk particles. Changes to the ECH toroidal angle therefore cause greater relative changes to the electron density at $\mathcal{E}=\qty{90}{keV}$. Specifically, at $t=\SI{1.05}{s}$, while the electron density at $\mathcal{E}=\qty{20}{keV}$ only rises by a factor of 3, the electron density at $\mathcal{E}=\qty{90}{keV}$ by a factor of 274. Considering that the VECE diagnostic requires a certain number of electrons to detect a signal, it is only after an additional rise in the local electron density at $\qty{90}{keV}$ by a factor of 12 that the VECE diagnostic detects the presence of superthermal electrons (c.f.\ figure~\ref{fig:VECE_synthetic_72644}). Additionally, for the \qty{96}{GHz} signal of discharge \#72644, the deviation observed in the final timestep is most likely due to the presence of a substantial runaway electron population, which is not accounted for in the \LUKE--\YODA\ calculation. This interpretation is supported by Hard X-ray measurements up to \qty{1.8}{\second}, where the blind channel~25 of camera~C5 records strong radiation from $t \approx \qty{1.5}{\second}$, indicative of runaway electrons.

\begin{figure}
    \centering
                \begin{overpic}[width=0.49\textwidth]{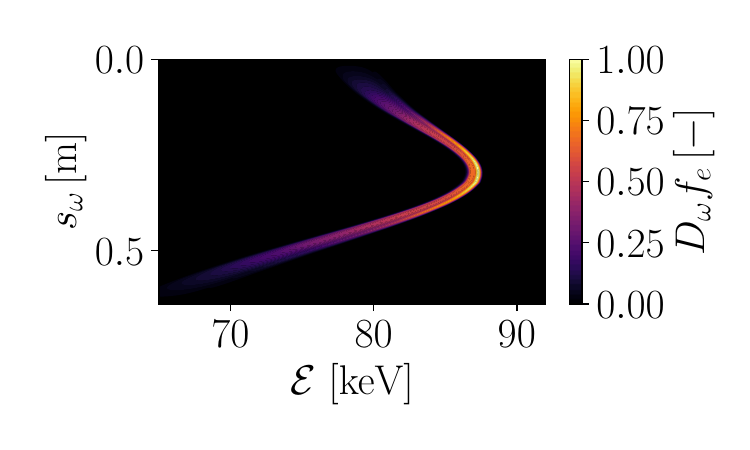}
        \put(22,54){(a) $t=\qty{1.05}{s}$}
    \end{overpic}
    \begin{overpic}[width=0.49\textwidth]{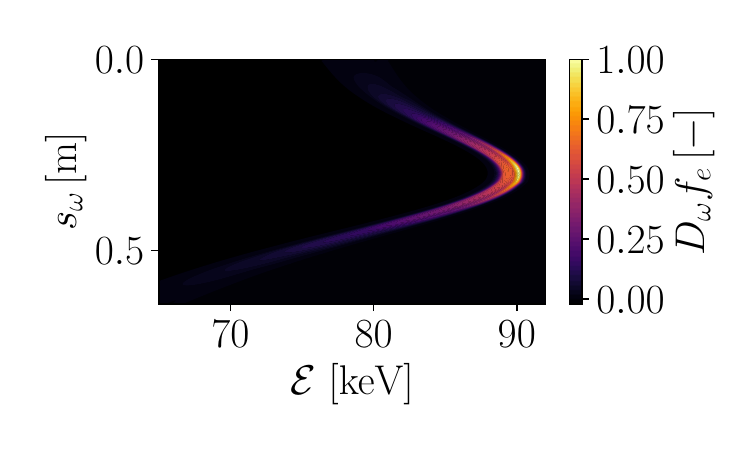}
        \put(22,54){(b) $t=\qty{1.65}{s}$}
    \end{overpic}
    \caption{Birthplace maps to the \(96.35\,\mathrm{GHz}\) VECE channel at (a) \(t=1.05\,\mathrm{s}\) and (b) \(t=1.65\,\mathrm{s}\). Plotted is the emissivity-weighted contribution density \(f\!\big(r(s_\omega),E\big)\,D_{\omega,e}(s_\omega,\mathcal{E})\) versus electron energy \(\mathcal{E}\) and position along the line of sight \(s_\omega\) ($s_\omega=0$ corresponds to the antenna position). Bright colors indicate larger contributions. In both cases the signal is dominated by electrons near the plasma core with \(\mathcal{E}\approx90\,\mathrm{keV}\); at \(t=1.65\,\mathrm{s}\) the contributing region is slightly broader and shifted outward. Color bars are normalized independently for each panel.}

    \label{fig:72644:green}
\end{figure}

To determine which parts of the distribution function contribute most strongly to the VECE signal, \YODA{} can be used to calculate the Green's function $D_\omega(s,p_\parallel)$. This function encodes the contribution to the VECE signal from different points $s$ along the line-of-sight, and from particles with different $p_\parallel$, and allows the determination of which features the detector is most sensitive to in the distribution function. The Green's function is defined via its relation to the signal intensity $I_\omega$ and $f$ through the integral
\begin{equation}\label{eq:green}
    I_\omega = \int\int
        f\left[r(s),p_\parallel\right]
        D_\omega\left(s,p_\parallel\right)\,
        \dd s\dd p_\parallel.
\end{equation}
The integrand of equation~\eqref{eq:green}, here referred to as the birthplace distribution $fD_\omega$, is shown in figure~\ref{fig:72644:green} for the \qty{96.35}{GHz} (\qty{90}{keV}) channel in the geometry of \#72644 at two different times during the discharge. As the ECH toroidal angle changes, so does the regions of the plasma which dominate the emission. At early times, the suprathermal particles are spread relatively evenly across the radial coordinate, whereas at later times the suprathermal population near the magnetic axis has grown significantly larger than at other radii.

The birthplace distribution is mapped in Fig.~\ref{fig:72644:green} as a function of electron energy \(\mathcal{E}\) and ray path \(s_\omega\). At \(t=1.05\,\mathrm{s}\) the contribution is already concentrated around the plasma core and peaks at \(\mathcal{E}\approx90\,\mathrm{keV}\), indicating that a nascent suprathermal tail governs the 96.35\,GHz signal.
By \(t=1.65\,\mathrm{s}\) the contributing region broadens, consistent with the slow evolution of the resonant surface and the growth of the high-energy tail seen in Fig.~\ref{fig:72644:f}.

%%%%%%%%%%%%%%%%%%%%%%%%%%%%%%%%%%%%%%%%%%%%%%%%%%%%%%%%%%%%%%%%%%%%%%%%%%%%%%%%%%%%%%%%%%%%%%%%%%%%%%%%%%%%%%%%%%%%%%%%%%%%%%%%%%%%%%%%%%%%%%%%%%%%%%%%%%%%%%%%%%%%%%%%%%%
%%%%%%%%%%%%%%%%%%%%%%%%%%%%%%%%%%%%%%%%%%%%%%%%%%%%%%%%%%%%%%%%%%%%%%%%%%%%%%%%%%%%%%%%%%%%%%%%%%%%%%%%%%%%%%%%%%%%%%%%%%%%%%%%%%%%%%%%%%%%%%%%%%%%%%%%%%%%%%%%%%%%%%%%%%%
\subsection{Time-dependent Fokker-Planck simulations}
When simulating fast transients, such as the ECH on/off phases or rapid power modulations \cite{Choi2019SuprathermalED,Cazabonne_2023}, the assumption of a
radially uniform loop voltage breaks down. In such cases, time-dependent simulations with \LUKE{} are necessary. The initial distribution function in these simulations is taken to be Maxwellian, which is then advanced in time through the Fokker-Planck equation.
Additionally, the electric field is evolved in time and space by solving the induction equation~\eqref{eq:LUKEinduction}, with the current density calculated from the evolving
distribution function.
 
Time-dependent simulations have been conducted for discharge \#73217, where the ECH power remained constant with a fixed injection angle for approximately 1 second before being
turned off. The simulation was run for a total duration of 1.8 seconds, using a time step of approximately 16 ms, corresponding to the time resolution of the Thomson
scattering diagnostic. In this model, electron radial transport was neglected (i.e.\ $\mathcal{D}_r = 0$). The electron-driven toroidal current calculated
by \LUKE{}, accounting for both ECCD and ohmic heating, matched the plasma current with less than 10\% discrepancy.

\begin{figure}
    \centering
    \includegraphics[width=0.8\textwidth]{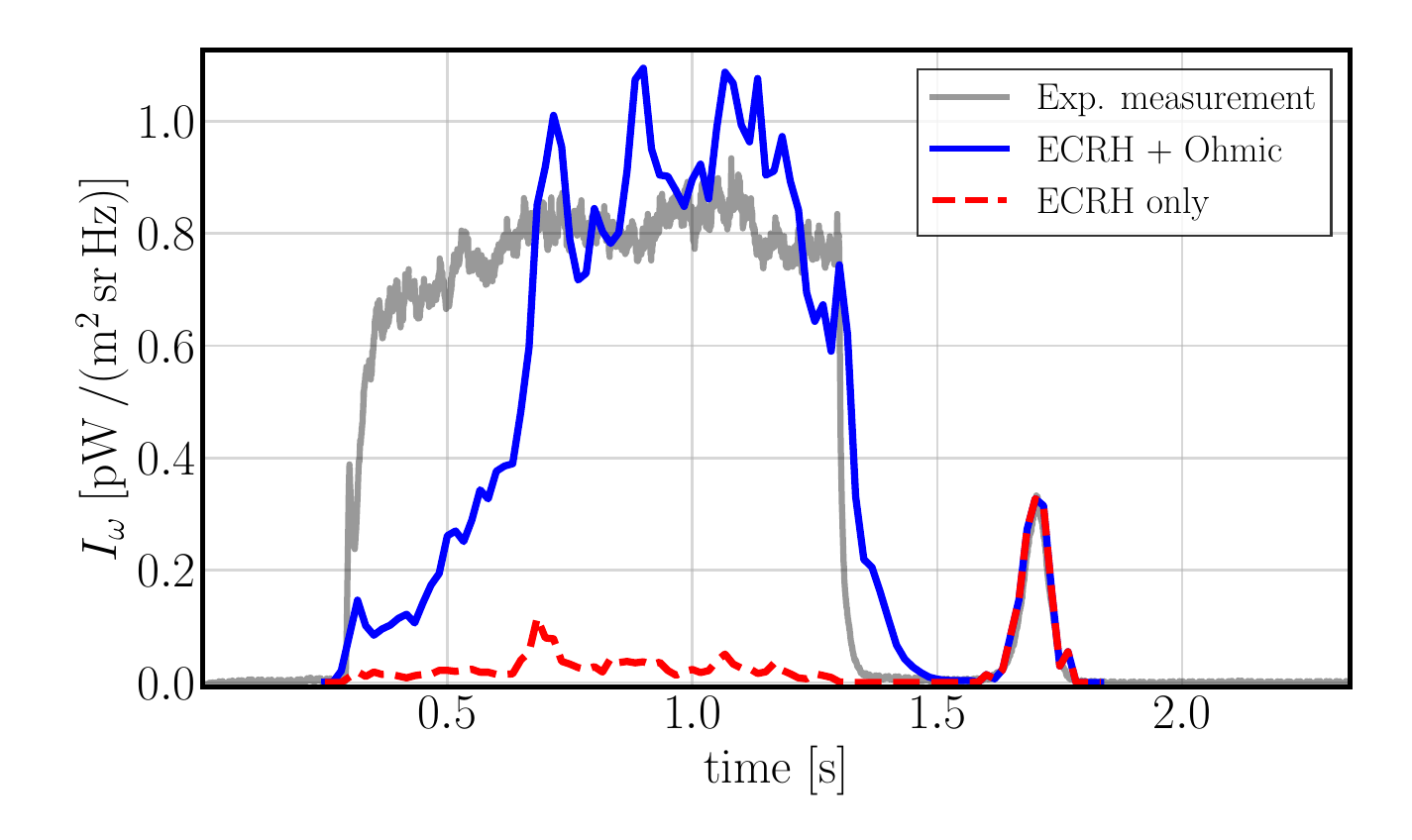}
    \caption{Comparison between VECE normalized raw data (solid gray line) and the modelled intensity (solid blue line) at frequency $f_{\rm ece}\sim\qty{109}{GHz}$ (3rd harmonic X-mode), corresponding to an electron kinetic energy of $ \mathcal{E} \sim \qty{80}{keV}$, in discharge \#73217. The red dashed line shows the VECE signal when the ohmic electric field is disabled, so that electron acceleration is due to ECRH only.}
    \label{fig:VECEmodel73217}
\end{figure}

As illustrated in Fig.~\ref{fig:VECEmodel73217}, at a frequency of approximately \SI{109}{GHz} (corresponding to an electron energy of \SI{80}{keV}), the model successfully reproduces several features of the VECE intensity. Notably, the model approximately reproduces the intensity at steady state, the decay of the signal when ECH is turned off, as well as the peak around $\SI{1.7}{s}$.

When the ECH is turned off at $t \sim\SI{1.3}{s}$ and the magnetic field remains constant for approximately \SI{100}{ms}, the model accurately predicts the drop in VECE 
intensity. This decrease occurs because ECH power is the primary driver of the fast electron population, and when it is no longer applied, the electron distribution function
thermalizes collisionally within a few milliseconds. The fact that the modeled VECE intensity does not immediately drop to zero after the ECH is turned off is attributed to the fact that the time step in \LUKE{} (\SI{16}{ms}) is longer than the decay time scale (few ms).

The peak around $t=\SI{1.7}{s}$ results from thermal ECE which is picked up when the magnetic field is ramped at the end of the discharge (c.f. figure~\ref{fig:exp73217}). This peak allows the diagnostic to be calibrated using the technique developed in Ref.~\cite{Arsene2024}, as the intensity of the peak should correspond to that emitted by a black-body at the instantaneous temperature. Hence, we normalize both the simulated and measured signal based on the value of this peak and find that not only are many of the dynamic features of the fast electron cyclotron emission reproduced, but so is the magnitude of the radiation as compared to the thermal emission. This suggests that not only the time dynamics are captured, but so is the number of fast electrons estimated by \LUKE{}, which adds further confidence to the validity of the model.

The thermal peak at $t=\qty{1.7}{ms}$ allows the diagnostic to be calibrated in one additional way. In the experimental setup, there is some uncertainty in the tilt of the field-of-view (FOV) relative to the vertical. If the FOV in the simulation is misaligned compared to the experimental view, it would either cross through regions of weaker or stronger magnetic field. Since the observation instant of the thermal peak is related to the time at which the magnetic field in the FOV is such that radiation is emitted at a frequency $f_{\rm ece}=\qty{109}{GHz}$, this means that the simulated peak would either occur slightly after or before the measured peak. In fact, through this technique it was found that the FOV of the VECE diagnostic was tilted by $\qty{0.5}{\degree}$.

The slow rise of the VECE signal is attributed to discrepancies in the simulated current profile. A careful examination of the simulation data reveals that the core loop voltage drops significantly below the edge value when ECRH is enabled, only slowly recovering during the ECRH phase before reaching an approximately radially uniform loop voltage at $t\approx\qty{0.75}{s}$, when the VECE signal first matches the experimental magnitude. This is hypothesized to be due to differences in the simulated and experimental current profiles, partially due to the EC wave deposition broadening and fast electron radial transport~\cite{Cazabonne_2023,Cazabonne_2024}. The physical details of this mechanism are not yet fully understood and are thus difficult to model, and so a careful investigation of the rise of the VECE signal in figure~\ref{fig:VECEmodel73217} is left for future study.

The interplay between the ohmic electric field and the injected EC waves is illustrated by the stark difference in figure~\ref{fig:VECEmodel73217} between the solid blue and dashed red lines. While the former includes a self-consistent solution of the induction equation with the experimentally measured edge loop voltage $V_{\rm loop}$ for boundary condition, the latter sets $V_{\rm loop}=0$ so that all electron acceleration comes from interactions with the EC waves only. The resulting VECE signal suggests that both the ohmic electric field and EC waves must be accounted for to accurately simulate the distribution function.

Another process which could influence the suprathermal electron dynamics and is not captured in our Fokker-Planck simulations are knock-on collisions, which in combination with the ohmic and ECRH acceleration could lead to a process similar to the avalanche multiplication of runaway electrons~\cite{Rosenbluth1997,Embreus2018}.
In this process, a suprathermal electron, initially accelerated by the EC wave interaction, can transfer a significant amount of energy to a thermal electron through a close Coulomb collision. This newly energized electron can then be further accelerated by the EC wave, entering the suprathermal regime and contributing to the overall growth of the fast electron population. Such an effect could be significantly enhanced by ohmic electric field acceleration even if the electron does not exceed the critical energy for runaway.

\section{Conclusions}
\label{sec:conclusions}
This study provides an investigation into suprathermal electron dynamics in the TCV tokamak measurements and simulation of the vertical ECE diagnostic. A new ECE synthetic diagnostic framework, \YODA, has been developed to simulate the emission and (re)absorption of electron cyclotron radiation for arbitrary electron distributions and antenna geometries. \YODA{} has been validated against the synthetic ECE code \SPECE{}, and the two codes agree well when applied to the TCV thermal plasma discharge \#73003. The 3D bounce-averaged relativistic Fokker-Planck code \LUKE{} has been employed to model electron distributions in two electron cyclotron current drive (ECCD) discharges (\#72644 and \#73217), which are compared against experimental VECE measurements. 

Time-asymptotic simulations of TCV discharge \#72644, in which the EC injection angle was varied to yield a staircase shaped VECE signal, agree well with the experimental measurements. By analyzing the \LUKE{} and \YODA{} simulations, it was found that the rise in VECE intensity as the EC toroidal angle is increased is due to the increase in electron density at the energy observed by each VECE channel. At \qty{20}{keV}, close to the thermal bulk, a tail of suprathermal electrons is quickly created already at small EC toroidal angles, and subsequent steps in angle only raise the density slightly. In contrast, at higher energies the electron tail remains very small until a substantial amount of EC power is deposited in the plasma, and only produces detectable levels of ECE at larger toroidal angles.

While the overall rise in the VECE signal is captured by the combined \LUKE{} and \YODA{} simulations, the relative change in signal level between the different EC toroidal angle steps is not always accurately captured. The exact cause for this remains unknown, but may be due to that the plasma does not have sufficient time to relax to a steady-state, as is assumed in the simulations. As noted with the time-dependent simulations, it is also possible that large-angle collisions could play a more important role in the dynamics than anticipated.

Time-dependent simulations of the distribution function accurately reproduce the decrease in VECE intensity when the ECH was turned off,  successfully capturing the rapid thermalization of the electron population. A key challenge, however, is the inability of the simulations to fully replicate the rapid increase in VECE  intensity when ECH power is first applied. This discrepancy suggests that additional physical processes, such as EC wave deposition broadening and fast electron radial transport, may play a role in the buildup of the suprathermal electron population.  Incorporating more advanced models to account for these potential effects could help improve the predictive capability of Fokker-Planck simulations in future studies and devices.

In conclusion, this work successfully demonstrates that the VECE diagnostic can serve as an effective additional experimental constraint to Fokker-Planck simulations, complementing other common diagnostics such as hard x-ray spectroscopy. The VECE diagnostic provides valuable insights into suprathermal electron dynamics, especially by offering detailed energy-resolved measurements that help refine the accuracy of simulated electron distributions.

With the framework presented here, it should now be possible to perform a detailed validation of the Fokker-Planck theory for suprathermal electrons. For example, with the help of specifically tailored experiments, it might be possible to validate the theory for Dreicer generation of runaway electrons~\cite{Connor1975}, as well as the effective critical electric field~\cite{Aleynikov2015,Stahl2015,Hesslow2018} for the same.

A natural next step in this line of research is to combine the use of VECE data with HXRS data, and other possible fast electron diagnostics, to more tightly constrain the distribution function. By incorporating the present framework with \LUKE{} and \YODA{} into a Bayesian statistics framework, such as described in~\cite{Jarvinen2022,Fischer2010}, could facilitate more accurate determinations of the fast electron distribution function.

\section*{Aknowledgements}
This work has been carried out within the framework of the EUROfusion Consortium, partially funded by the European Union via the Euratom Research and Training Programme (Grant Agreement No 101052200 — EUROfusion). The Swiss contribution to this work has been funded by the Swiss State Secretariat for Education, Research and Innovation (SERI). Views and opinions expressed are however those of the author(s) only and do not necessarily reflect those of the European Union, the European Commission or SERI. Neither the European Union nor the European Commission nor SERI can be held responsible for them. This work was supported by the Swedish Research Council (Dnr.\ 2024-04879).

\section*{References}

\bibliography{biblio}

\appendix
\section{Derivation of the induction equation in toroidal geometry}
\label{AppendixA}
The general form of the induction equation is obtained by combining Ampère's and Faraday's laws: 
\begin{equation}
\label{eq:Ampere}
    \nabla \times \boldsymbol{B}=\mu_0 \boldsymbol{J},
\end{equation}
\begin{equation}
\label{eq:Faraday}
    \nabla \times \boldsymbol{E}=-\frac{\partial \boldsymbol{B}}{\partial t},
\end{equation}
Taking the curl of both equations and using the relations $\nabla \times (\nabla \times \boldsymbol{E})=-\nabla^2 \boldsymbol{E}$ and $\nabla \cdot \boldsymbol{E}=0$, we obtain the general form of the induction equation
\begin{equation}
\label{eq:inductionApp}
    \nabla^2 \boldsymbol{E}=\mu_0\frac{\partial \boldsymbol{J}}{\partial t}.
\end{equation}
Considering only the toroidal component of this equation and using $\boldsymbol{E}=E_\phi(\psi,\theta)\hat{\phi}$, which satisfies $\nabla \cdot \boldsymbol{E}=0$, the projection of (\ref{eq:inductionApp}) in the toroidal direction gives 
\begin{equation}
    \hat{\phi}\cdot \nabla^2(E_\phi(\psi,\theta)\hat{\phi})=\mu_0 \frac{\partial J_\phi}{\partial t}.
\end{equation}
Taking into account that $\nabla^2 ({E}_\phi \hat{\phi})= -\nabla \times (\nabla \times {E}_\phi \hat{\phi})$, the induction equation \ref{eq:inductionApp} can be written as
\begin{equation}
\label{eq:inducA5}
    \hat{\phi} \cdot \nabla \times ({E}_\phi \nabla \times \hat{\phi} - \hat{\phi} \times \nabla {E}_\phi) + \mu_0 \frac{\partial J_\phi}{\partial t} = 0.
\end{equation}
Now we introduce the $(\psi,s,\phi)$ coordinates, where $\psi$ is the poloidal flux, $s$ the arc length along a flux surface in the poloidal direction, and $\phi$ the toroidal angle. This set is used to parametrize closed flux-surfaces, and is defined from the origin $(R_p, Z_p)$ on the (closed) space
\begin{eqnarray}
                \psi\in\left[\psi_0,\psi_a\right], \quad
    s\in\left[0, s_{\rm max}\right]
\end{eqnarray}
which is related to $(r,\theta,\phi)$ by $\psi=\psi(r,\theta)$ and $s=s(r,\theta)$ \cite{LUKE3Dkinetic}. By expressing the gradient and curl operators in the $(\psi,s,\phi)$ coordinate system,
it is possible to rewrite the argument of the first curl in equation \ref{eq:inducA5} as
\begin{equation}
    E_{\phi} \nabla \times \hat{\phi} - \hat{\phi} \times \nabla E_{\phi}
= \hat{\psi} \frac{1}{R} \frac{\partial (RE_{\phi})}{\partial s} - \hat{s} |\nabla \psi| \frac{\partial (RE_{\phi})}{\partial \psi},
\end{equation}
so that the left hand side of the induction equation becomes
\begin{eqnarray}
\label{eq:indA7}
\nonumber
    \left(\hat{\phi} \cdot \nabla\right) \mathbf{E} &=
    -|\nabla \psi| \frac{\partial}{\partial \psi} \left(
        \frac{|\nabla \psi|}{R}
        \frac{\partial (RE_{\phi})}{\partial \psi}
    \right) - \\
    &- |\nabla \psi| \frac{\partial}{\partial s} \left(
        \frac{1}{R|\nabla \psi|}
        \frac{\partial (RE_{\phi})}{\partial s}
    \right)
    = -\frac{\partial J_{\phi}}{\partial t}. 
\end{eqnarray}

To further simplify the equation above, we introduce the flux surface average of a physical quantity $\Gamma(\psi,s)$ as
\begin{eqnarray}
    \langle \Gamma \rangle(\psi) = \frac{1}{\bar{q}} \int_{0}^{s_{\text{max}}} \dd s \, \frac{B_0}{ 2\pi |\nabla \psi|} \Gamma(\psi, s),
\end{eqnarray}
where 
\begin{equation}
    \bar{q}(\psi) = \int_{0}^{s_{\mathrm{max}}} \dd s\, \frac{B_0}{2 \pi |\nabla \psi|}.
\end{equation}
Applying this operation to the equation \ref{eq:indA7}, we find that the $\partial/\partial s$ vanishes identically, and we are left with the equation
\begin{equation}
\frac{B_0}{\bar{q}}\frac{\partial}{\partial \psi}\left[ \int_0^{\mathrm{s_{max}}} \dd s \, \frac{|\nabla \psi|}{2 \pi R}\frac{\partial (RE_\phi)}{\partial \psi}\right] = \mu_0 \frac{\partial \langle J_\phi \rangle}{\partial t}.
\end{equation}
Since $RE_\phi$ is constant on a flux surface, it can be moved outside of the integral, and we thus introduce the function
\begin{equation}
    l(\psi) = \int_0^{\mathrm{s_{max}}} \mathrm{d} s \, \frac{|\nabla \psi|}{2 \pi R}=\frac{1}{B_0}\left \langle \frac{|\nabla \psi|^2}{R} \right \rangle.
 \end{equation}
 Finally, by introducing the loop voltage as $V_l=2\pi RE_\phi$ we obtain the \LUKE{} form of the induction equation
 \begin{equation}
     \frac{B_0}{\bar{q}}\frac{\partial}{\partial \psi}\left[B_0 l(\psi)\frac{\partial V_l}{\partial \psi} \right]= 2\pi\mu_0 \frac{\partial \langle J_\phi \rangle}{\partial t}. 
 \end{equation}

\end{document}